\documentclass[10pt,letterpaper,superscriptaddress]{article}
\usepackage{opex3}
\usepackage{graphicx}
\usepackage{color}

\begin{document}
\title{Spectral properties of high-gain parametric down-conversion}

\author{K.~Yu.~Spasibko$^1$, T.~Sh.~Iskhakov$^{2}$, M.~V.~Chekhova$^{2,1}$}
\address{$^1$Department of Physics, M.~V.~Lomonosov Moscow State University,
Leninskie Gory, 119992 Moscow, Russia \\
$^2$Max Planck Institute for the Science of Light,
G\"unther-Scharowsky-Stra\ss{}e 1/Bau 24, 91058 Erlangen, Germany} \email{drquantum@hotmail.com}

\begin{abstract}
High-gain parametric down-conversion (PDC) is a source of bright squeezed vacuum, which is a macroscopic nonclassical state of light and a promising candidate for quantum information applications. Here we study its properties, such as the intensity spectral width and the spectral width of pairwise correlations.
\end{abstract}
\ocis{(270.0270) Quantum optics; (270.5585) Quantum information and
processing}

Bright squeezed vacuum (BSV), the state generated via high-gain parametric down-conversion (PDC) and having a large number of photons per mode, is currently in the focus of attention in quantum optics and quantum information. The reason is that BSV is considered as a state that can manifest macroscopic entanglement~\cite{entanglement}. At the same time, because BSV can have huge numbers of photons per mode, it can be easily involved into nonlinear optical interactions~\cite{diTrapani} and interactions with material quantum objects such as atoms, molecules, quantum dots etc. In particular, its application to quantum memory is discussed~\cite{Kupriyanov}. Recently, BSV has been used for quantum imaging~\cite{Brida_Nature} and absolute calibration of photodetectors~\cite{calibration}. Because of these applications, it is very important to know the spectral properties of BSV emitted via high-gain PDC.

It should be noted here that the spectra of PDC are essentially two-dimensional (see, for instance, Ref.~\cite{Big}): the shape of the frequency spectrum depends on the selected angle of emission, and vice versa. In this connection, it is worth mentioning an interesting concept of two-dimensional coherence of PDC radiation~\cite{Xcoh}. Therefore, a general consideration should involve both frequency and angular spectra. However, in this work we only consider frequency spectra and all measurements are performed with an aperture selecting only collinear emission.

At high gain, PDC does not any more generate two-photon light, as is the case of low gain. The radiation of high-gain PDC contains not only photon pairs, but four-photon states, six-photon states, and so on, and the mean photon number per mode is large. The mean photon number per mode is in one-to-one correspondence with \textit{brightness}; this is why the state can be called \textit{bright squeezed vacuum}. Under these conditions, correlations in SV cannot be efficiently studied by measuring the number of coincidences. Instead, one should measure the noise (variance) of the intensity difference for the two selected beams~\cite{Jedrkiewicz,Bondani,JETPLett}.

The spectral and angular properties of high-gain PDC (known earlier as parametric super-fluorescence~\cite{DNK1}) have been well studied theoretically~\cite{DNK1,Brambilla,Dayan}. In particular, it has been known for a long time~\cite{DNK1} that the spectral width of both signal and idler beams should increase with the gain. Physically, this is clear from the fact that at high gain, there is exponential amplification of down-converted light along the crystal. Therefore, effectively it is not the whole crystal that contributes into the signal and idler intensity but only its exit part, where the signal/idler intensities (photon numbers) are most high. For the same reasons, the spectral width of correlations, usually determined by the inverse pump pulse width/coherence time, should also increase at high gain. Similar behavior should be observed for the angular spectra: both the total angular width and the far-field zone speckle size (correlation radius) should increase at high gain. Indirect experimental evidence for this was observed in Ref.~\cite{Jedrkiewicz}. The speckle size has been directly measured as a function of the parametric gain in Ref.~\cite{Ivano}. However, no measurements on the frequency spectra of high-gain PDC have been performed, although there was an experiment on the study of spectral correlations for a similar effect, four-wave mixing~\cite{Kerr}.

In this work, we study the spectral properties of BSV generated via high-gain PDC from a picosecond-pulse pump. Due to a rather short coherence time of the pump, we are able to resolve the spectral width of correlations between signal and idler radiation. By scanning both the signal and idler frequencies separately, we can see the two-dimensional distribution of correlations, similar to the 'joint two-photon intensity' observed for two-photon light~\cite{Kim&Grice,Poh,Kim}. We also measure the spectral width versus the parametric gain and observe the spectral broadening in accordance with~\cite{DNK1}. Finally, we observe some interesting effects caused by super-bunching in high-gain PDC.

\begin{figure}
\begin{center}
\includegraphics[width=0.7\textwidth]{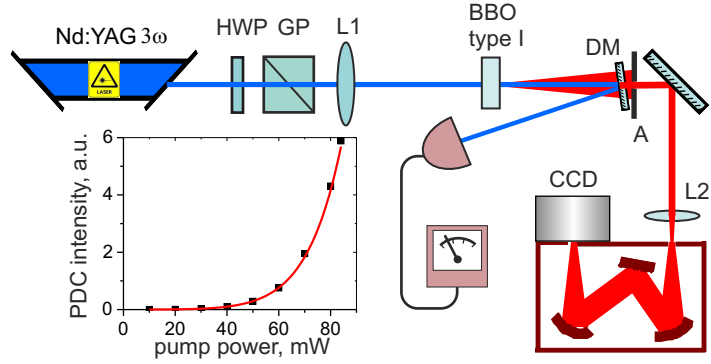}
\caption{(color online) The experimental setup. PDC is generated in a type-I BBO crystal with the 3rd harmonic of Nd:YAG laser as a pump. The pump power can be changed using a halfwave plate HWP and a Glan prism GP. The pump is eliminated with a dichroic mirror DM. For PDC radiation, a narrow angular spectrum is selected with an aperture A. The spectral properties are studied using a spectrometer and a CCD camera. The inset shows the dependence of the PDC output intensity on the pump power.}
\label{setup}
\end{center}
\end{figure}

The experimental setup is shown in Fig.~\ref{setup}. BSV was obtained in a $2$ mm long BBO crystal cut for type-I collinear frequency-degenerate phasematching. As the pump, we used the 3rd harmonic of a Nd:YAG laser with the wavelength $\lambda_p=354.7$ nm, pulse duration $18$ ps, repetition rate $1$ kHz and the energy per pulse up to $0.1$ mJ. The coherence time of the pump was less than the pulse duration and amounted to $5$ ps. To increase the parametric gain, the pump was focused into the crystal by means of a lens (L1) with the focal length $1$ m. This resulted in the maximal peak intensity $10$ GW/cm$^2$. The gain value could be varied by changing the pump power, which was done by rotating a half-wave plate (HWP) in front of a Glan prism (GP). After the crystal, the pump radiation was reflected by a dichroic mirror (DM), which transmitted nearly all the down-converted light (97\%), and its power was measured with a power-meter. The PDC radiation was filtered in the transverse wavevector (angle of emission) by means of an aperture (A) placed in the far-field zone. The emission angle selected this way was $0.17^{\circ}$ (inside the crystal). All radiation passing through the aperture was focused, using a lens (L2), on the input slit of the HORIBA Jobin Yvon Micro HR monochromator equipped with a SYGNATURE PDA CCD array. The spectral measurements were performed with the best possible resolution, which was $0.2$ nm.

In all our measurements, the spectral properties of PDC were studied depending on the parametric gain values. In its turn, the parametric gain was found by fitting the dependence of the PDC output intensity $I$ on the input pump power $P$ (inset in Fig.~\ref{setup}) by the function
\begin{equation}
I=I_0\sinh^2(\kappa\sqrt{P}),
 \label{gain}
\end{equation}
where $I_0,\kappa$ were the fitting parameters~\cite{Ivanova}. The gain was measured for collinear frequency-degenerate regime. It is important that for such a measurement, the PDC radiation should be very well filtered both in the angle and in frequency, so that less than a single mode is registered. After the fitting, each pump power is put into correspondence to a certain value of the parametric gain $G\equiv\kappa \sqrt{P}$.

\begin{figure}
\begin{center}
\includegraphics[width=0.6\textwidth]{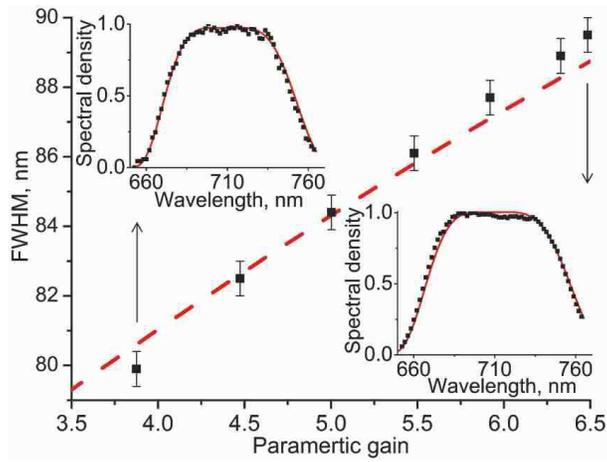}
\caption{(color online) The dependence of the PDC spectral width on the parametric gain. The insets show the shapes of the spectra at gain values $3.9$ (left) and $6.5$ (right). The dashed line is exact calculation using the theoretical approach of Ref.~\cite{Dayan}.}
\label{width}
\end{center}
\end{figure}

The dependence of the PDC spectral width (FWHM) on the parametric gain was measured for the range of gain values  $3.9<G<6.5$. At lower gain values, the PDC signal was too low to make spectral measurements while higher gain values could not be achieved with the available pump powers. The measured dependence is shown in Fig.~\ref{width}. One can see that in the chosen range of gain values, the spectral width changes by 12\%. Compared to the spectral width of low-gain PDC, the increase is $27\%$. Note that the effect of the gain on the spectral width is even stronger in the case of type-II or frequency-nondegenerate phase matching~\cite{DNK1}. The insets show the shapes of the spectra recorded at the lowest and highest gain values. The shapes are well fitted with the theoretical dependencies calculated using the approach of Ref.~\cite{Dayan}. The same approach was used in calculating the theoretical dependence of the FWHM on the parametric gain (dashed line). One can observe perfect agreement between the theory and experiment. The only fitting parameter used in the theoretical dependence was the orientation of the crystal, which turned out to differ by $0.0025^{\circ}$ from the one for exact collinear degenerate phase matching. This led to about $2$ nm shift for the whole dependence, which was difficult to notice in the experiment.

\begin{figure}
\begin{center}
\includegraphics[width=0.4\textwidth]{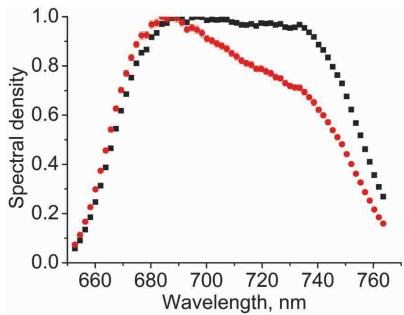}
\caption{(color online) The measured type-I PDC spectrum before (red circles) and after (black squares) the correction by $\lambda^{4}$ factor. The spectrum corresponds to the gain value $G=6.48\pm0.05$.}
\label{spectrum}
\end{center}
\end{figure}

Here, it is worth mentioning that the `raw' spectra obtained at the output of the spectrometer were always asymmetric, the long-wavelength part having smaller intensity. A typical distribution is shown in Fig.~\ref{spectrum} by red circles. It is interesting that similar distributions have been reported in many papers on degenerate (broadband) PDC - see, for instance, Ref.~\cite{Kim}.

By analyzing the detection scheme (standard for such measurements), which includes the selection of a certain small solid angle and a certain wavelength interval, we come to the conclusion that the spectra should be multiplied by a factor scaling as the fourth degree of the wavelength, $\lambda^4$. The reason is that the total number of photons is the product of the number of photons per mode (wavelength-independent in the vicinity of the degenerate wavelength) and the number of modes collected. At the same time, the number of transverse and longitudinal modes within a given solid angle interval $\Delta\Omega$ and a given wavelength interval $\Delta\lambda$ scales as $1/\lambda^4$. This can be understood from the following considerations. The number of modes is given by the ratio of the detection volume and the coherence volume~\cite{DNK},
\begin{equation}
m=\frac{V_{det}}{V_{coh}}.
 \label{m}
\end{equation}
 The coherence volume is $V_{coh}=S_{coh}ct_{coh}$, where $S_{coh}$ is the transverse coherence area, $c$ is the speed of light, and $t_{coh}$ is the coherence time. The detection volume does not depend on the wavelength, while the coherence area and time are wavelength-dependent,
 \begin{equation}
S_{coh}\sim \frac{1}{(\Delta k_{\bot})^2}\sim\lambda^2,\,\,\,t_{coh}\sim\frac{1}{\Delta\omega}\sim\frac{\lambda^2}{\Delta\lambda},
 \label{coh}
\end{equation}
where $\Delta k_{\bot}$ is the interval of transverse wavevector variation and $\Delta\omega$ is the frequency bandwidth
corresponding to $\Delta\lambda$. As a result, the number of collected modes scales as $\lambda^{-4}$, and this
factor should be taken into account in the spectra. Figure~\ref{spectrum} shows the shape of the spectrum before
(red circles) and after (black squares) the correction. The correction also took into account the fact that in our CCD camera, the wavelength interval selected by the one pixel slightly depended on the wavelength. This lead to a 10\% variation of the wavelength resolution along the CCD camera.

\begin{figure}
\begin{center}
\includegraphics[width=0.7\textwidth]{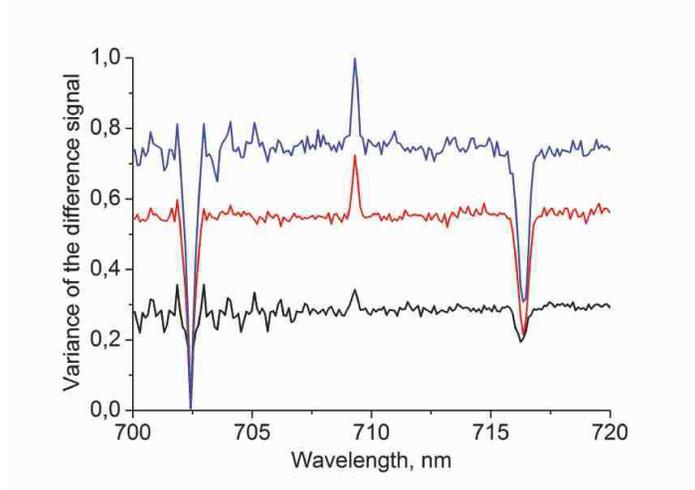}
\caption{(color online) Normalized variance of the difference between the PDC intensities at a given wavelength $702.4$ nm and scanned wavelengths. Different curves correspond to different gain values: 6.4 (lower curve, black), 6.5 (middle curve, red), and 6.8 (upper curve, blue).}
\label{correlations}
\end{center}
\end{figure}

In order to study the spectral correlations in high-gain PDC, we used the technique of noise reduction measurement~\cite{Jedrkiewicz,Bondani,JETPLett}, in which the observable under study is the variance of the difference of intensities for two beams. Correlation between two beams leads to a decrease in the noise of the difference intensity. We have studied the difference-intensity noise between beams at various pairs of wavelengths in the PDC spectrum. For different gain values, $10000 - 20000$ frames were made, each one collecting the intensity from $500-1000$ pulses, and then processed.  First, one wavelength was fixed (by fixing the pixel of the CCD camera) and the other one was scanned. The variance was calculated for the difference between the signals from the chosen pixel and all remaining ones. The results are shown in Fig.~\ref{correlations}. In all spectra, the fixed wavelength (`signal') was chosen to be $\lambda_s=702.4$ nm, while the other one (`idler'), $\lambda_i$, was scanned within the range from $700$ nm to $720$ nm.

One can see that each spectral distribution of the variance contains two `dips' and one peak. The left-hand `dip' is caused by the classical correlations existing within a single mode of PDC radiation. The dip goes down to zero because the variance is identically zero for the case where the reading of a pixel is subtracted from itself. This point is an artifact and should not be considered. Another artifact is the oscillating noise seen on some (but not all) spectra in the vicinity of the left-hand 'dip'. The right-hand dip is caused by the nonclassical correlation between light beams scattered into parametrically conjugated modes, with the wavelengths related as $\lambda_s^{-1}+\lambda_i^{-1}=\lambda_p^{-1}$.

%In theory, if the mean signals in the two pixels are equal, the variance value in this dip should be given
%by $(1-\eta)S$, where $\eta$ is the detection efficiency and $S$ is the mean signal of the pixel. In reality
%(see Fig.~\ref{spectrum}), the signals are different even after the correction by the factor $\lambda^{4}$.
%Therefore, the value of NRF in the `dip' is higher.

The peak in Fig.~\ref{correlations} occurs at exactly the degenerate wavelength $709.3$ nm. It appears because at the degenerate wavelength, PDC radiation is single-mode squeezed vacuum, sometimes referred to as 'light with even photon numbers', and this kind of radiation has peculiar statistical properties. Namely, single-mode squeezed vacuum  manifests super-bunching even at strong parametric gain~\cite{single-mode}. Its second-order normalized Glauber's correlation function is $g^{(2)}=3+1/N$, where $N$ is the mean photon number per mode, and in the high-gain regime $g^{(2)}=3$. Therefore, it has increased fluctuations, which causes the peak in Fig.~\ref{correlations}.

The dips and the peak in Fig.~\ref{correlations} can be explained in more detail by recalling that the variance of the photon-number difference between the beams labeled $s$ and $i$ is~\cite{STELLA}
\begin{equation}
\hbox{Var}(N_s-N_i)=(g^{(2)}_{ss}-1)\langle N_s\rangle^2+(g^{(2)}_{ii}-1)\langle N_i\rangle^2-2(g_{si}^{(2)}-1)\langle N_s\rangle\langle N_i\rangle+\langle N_s+N_i\rangle,
 \label{var}
\end{equation}
 where $g^{(2)}_{ss}, g^{(2)}_{ii}$ are second-order normalized correlation functions for the signal and idler beams and $g^{(2)}_{si}$ is their cross-correlation function. The last term in (\ref{var}) represents the shot noise and is negligible in the high-gain regime. Note that Eq. (\ref{var}), initially written for a single radiation mode, will not change if we assume that $m$ independent modes are registered: the variance in the left-hand side, as well as the mean values in the right-hand side, will acquire a factor of $m$, and the factors containing the correlation functions in the right-hand side will change as $g^{(2)}-1\rightarrow (g^{(2)}-1)/m$~\cite{Ivanova}.

 From the second term we see that the variance of the difference signal should have a peak if $\lambda_i$ corresponds to the degeneracy. Indeed, if $\langle N_s\rangle=\langle N_i\rangle\equiv N$, then outside of the peak, where $g^{(2)}_{ii}=g^{(2)}_{ss}=2$, the variance of the difference photon number is $2N^2$ while in the peak, $g^{(2)}_{ii}=3$, and the variance is $3N^2$. In the `autocorrelation dip', $\lambda_i\approx\lambda_s$,  $g^{(2)}_{ii}=g^{(2)}_{ss}=2$, and $\hbox{Var}(N_s-N_i)\approx2N$, i.e., the variance of the difference photon number should be suppressed down to the shot-noise level. In the `cross-correlation dip', where $\lambda_s$ and $\lambda_i$ are conjugated, $g^{(2)}_{ij}=2+1/N$, and in theory, $\hbox{Var}(N_s-N_i)$ should be zero. However, this complete suppression of noise can be observed only provided that three conditions are satisfied~\cite{JETPLett}: (1) the quantum efficiency is high; (2) a large number of transverse and longitudinal modes are collected and (3) the signal and idler modes are properly matched. In our experiment, the quantum efficiency is not more than 20\%, only a few  transverse modes are selected, and no proper mode matching is performed.

We see that with the increase of the gain, the `dips' in Fig.~\ref{correlations} become more pronounced. This is because the background is caused by the excess noise in the PDC intensity (the first two terms in (\ref{var})), which scales quadratically with the mean photon number. At the same time, the minimum in the `dip' is determined by the shot noise (last term in (\ref{var})), scaling linearly with the mean photon number. As the gain grows, the mean photon number increases and the `dip' becomes more visible.

Another observation one can make from Fig.~\ref{var} is that the peak width is considerably less than the width of the `dips'.
Accurate fitting of all of them by Gaussian functions yields, in the case of the middle dependence in
Fig.~\ref{correlations}, the values of $0.26\pm0.06$ nm, $0.45\pm 0.06$ nm and $0.52\pm0.06$ nm for the peak,
the `autocorrelation dip' and the `cross-correlation dip', respectively.
%The peak width is considerably influenced by the finite resolution of the spectrometer, while the two `dips' are
%broadened due to the finite resolution by only about $10\%$.
The widths of the dips coincide with the widths of the second-order correlation functions and indicate the spectra of frequency correlations.
Theoretically, they should be given by the pump spectral width, which is $\Delta\omega_p=1.25$ THz. This corresponds to $\Delta\lambda=0.28$ nm at wavelengths around $709$ nm, but at high gain should be larger due to the nonlinear dependence of the PDC intensity on the pump intensity.
%In particular, at the gain values 6.4-6.8 (Fig.~\ref{correlations}) the correlation width should be
%approximately $0.48$ nm.

%The dips become slightly broadened with the increase in the gain.
%This behavior is in agreement with the spectrum broadening, as well
%as with the results of Ref.~\cite{Ivano} on the speckle size dependence on the gain.

As to the peak, its width, theoretically, should be narrower than the pump width by a factor of $2$. The observed widths of $0.26\pm0.06$ nm for the peak and $0.52\pm0.06$ nm for the dip are in agreement with this figure.

\begin{figure}
\begin{center}
\includegraphics[width=0.6\textwidth]{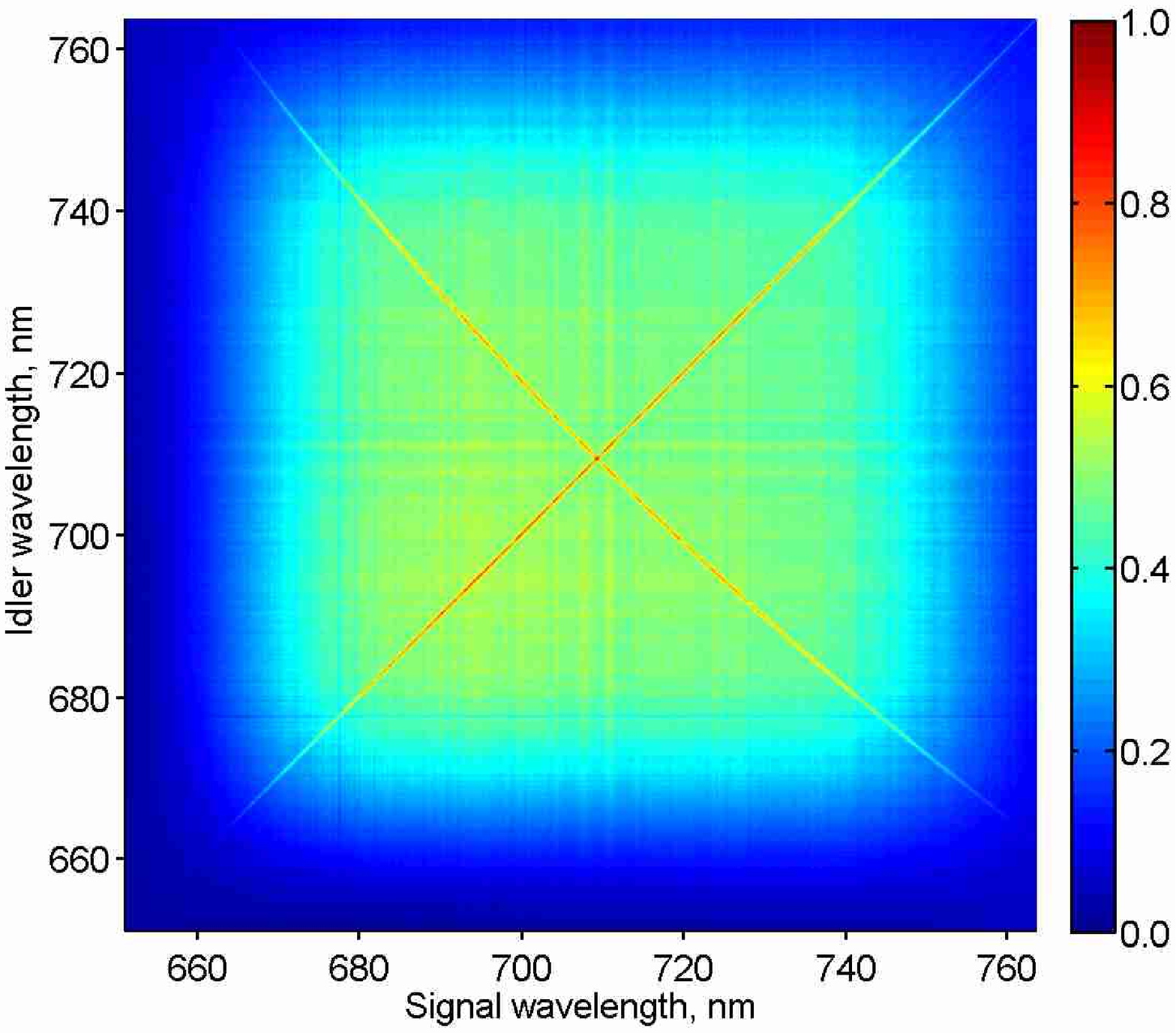}
\includegraphics[width=0.39\textwidth]{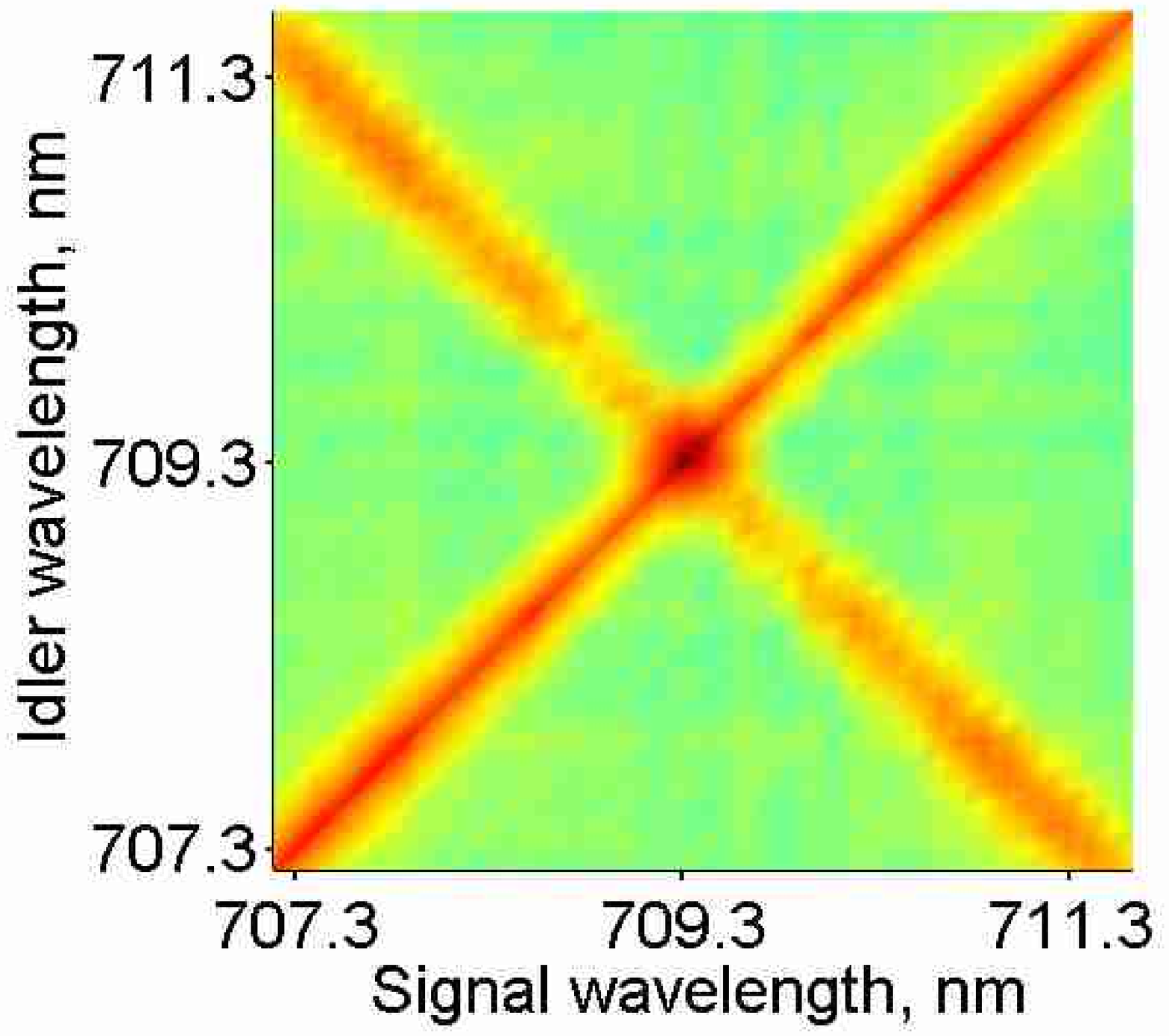}
\caption{(color online) Two-dimensional distribution of the covariance for the PDC intensities measured at signal and idler wavelengths (left) and its zoomed central part (right). The covariance is normalized to its maximum value.}
\label{2Dcor}
\end{center}
\end{figure}

More complete information about the frequency correlations in the spectrum of high-gain PDC can be obtained by
measuring a two-dimensional distribution of some parameter characterizing the correlations, as a function of both wavelengths $\lambda_s,\lambda_i$. As such a parameter, one can choose the difference-signal variance given by Eq.~(\ref{var}) but it is more convenient to measure the covariance of the signals at two various wavelengths, \begin{equation}
\hbox{Cov}(N_s,N_i)\equiv\langle N_s N_i\rangle-\langle N_s\rangle\langle N_i\rangle.
 \label{covar}
\end{equation}
Note that the same strategy was used in Ref.~\cite{Kerr}, where the correlation coefficient (normalized covariance) was measured to characterize intensity correlations in the spectra of bright beams.

The covariance is equal, up to a constant factor, to the third term in (\ref{var}), since
$(g_{si}^{(2)}-1)\langle N_s\rangle\langle N_i\rangle=\hbox{Cov}(N_s,N_i)$. It gives a full account of the pairwise correlations,
which now manifest themselves as peaks (Fig.~\ref{2Dcor}).
The `autocorrelation' peaks occur on the straight line $\lambda_s=\lambda_i$ and correspond to the classical correlations within a single PDC mode. The `cross-correlation' peaks correspond to the signal-idler correlations and are on the line given by $\lambda_s^{-1}+\lambda_i^{-1}=\lambda_p^{-1}$. The distribution of these peaks is similar to the 'joint intensity' observed for two-photon light~\cite{Kim&Grice,Poh,Kim}: its cross-section (conditional distribution) shows the spectral range of correlations while its projections on the axes $\lambda_s,\lambda_i$ (marginal distributions) show the PDC spectrum. The increased value of the correlation function for the collinear frequency-degenerate PDC also reveals itself in the vicinity of the point $\lambda_s=\lambda_i=709.3$ nm.

In conclusion, we have performed an experiment aimed at the study of the spectral properties of high-gain PDC. In agreement with the theory, we observed the broadening of the spectrum with the increase in the parametric gain. We have shown that the dependence of the coherence volume on the wavelength causes the asymmetry of the measured PDC spectra, which can be removed by multiplying the spectra by a factor of $\lambda^4$. By measuring the variance of the intensity difference for two wavelengths, we saw the correlations in the PDC spectrum: both auto-correlations, due to the thermal statistics of each mode of the PDC radiation, and cross-correlations, due to the pairwise statistics of signal and idler modes. The technique of noise reduction measurement allows us to resolve the spectral width of the intensity correlations. The measured  variance distributions also show the super-bunching of PDC radiation at exactly degenerate wavelength. Finally, we have measured the distribution of the covariance for the PDC intensities at two wavelengths, both scanned within the range $650-760$ nm. The obtained two-dimensional distributions show both the intensity correlations within a single mode of PDC radiation and the cross-correlation between the intensities at the signal and idler wavelengths.

We would like to thank Gerd Leuchs for illuminating discussions. This work has been supported in part by the Russian Foundation for Basic Research, grants 10-02-00202 and 11-02-01074. T.Sh.I. acknowledges funding from Alexander von Humboldt Foundation.

\end{document}